\documentclass[letterpaper,english]{emulateapj}
\usepackage[T1]{fontenc}
\usepackage[latin9]{inputenc}
\usepackage{amssymb}
\usepackage{graphicx}

\makeatletter

\newcommand{\source}{Aql~X-1}
\shorttitle{NICER Observes X-Ray Burst from Aql X-1}
\shortauthors{Keek, Arzoumanian, Bult, Cackett, et al.}

\makeatother

\usepackage{babel}
\begin{document}

\title{\emph{NICER} Observes the Effects of an X-Ray Burst on the Accretion
Environment in Aql~X-1}

\author{L.~Keek,\altaffilmark{1} Z.~Arzoumanian,\altaffilmark{2} P.~Bult,\altaffilmark{2}
E.\,M.~Cackett,\altaffilmark{3} D.~Chakrabarty,\altaffilmark{4}
J.~Chenevez,\altaffilmark{5} A.\,C.~Fabian,\altaffilmark{6} K.\,C.~Gendreau,\altaffilmark{2}
S.~Guillot,\altaffilmark{7,8,9} T.~Güver,\altaffilmark{10,11}
J.~Homan,\altaffilmark{12,13} G.\,K.~Jaisawal,\altaffilmark{5}
F.\,K.~Lamb,\altaffilmark{14,15} R.\,M.~Ludlam,\altaffilmark{16}
S.~Mahmoodifar,\altaffilmark{2} C.\,B.~Markwardt,\altaffilmark{2}
J.\,M.~Miller,\altaffilmark{16} G.~Prigozhin,\altaffilmark{4}
Y.~Soong,\altaffilmark{1,2} T.\,E.~Strohmayer,\altaffilmark{2}
M.\,T.~Wolff\altaffilmark{17}}

\altaffiltext{1}{Department of Astronomy, University of Maryland, College Park, MD 20742, USA}
\altaffiltext{2}{X-ray Astrophysics Laboratory, Astrophysics Science Division, NASA/GSFC, Greenbelt, MD 20771, USA}
\altaffiltext{3}{Department of Physics \& Astronomy, Wayne State University, 666 W. Hancock Street, Detroit, MI 48201, USA}
\altaffiltext{4}{MIT Kavli Institute for Astrophysics and Space Research, Massachusetts Institute of Technology, Cambridge, MA 02139, USA}
\altaffiltext{5}{National Space Institute, Technical University of Denmark, Elektrovej 327-328, DK-2800 Lyngby, Denmark}
\altaffiltext{6}{Institute of Astronomy, Madingley Road, Cambridge CB3 0HA, UK}
\altaffiltext{7}{Instituto de Astrofísica, Facultad de Física, Pontificia Universidad Católica de Chile, Av. Vicuña Mackenna 4860, 782-0436 Macul, Santiago, Chile}
\altaffiltext{8}{CNRS, IRAP, 9 avenue du Colonel Roche, BP 44346, F-31028 Toulouse Cedex 4, France}
\altaffiltext{9}{Université de Toulouse, CNES, UPS-OMP, F-31028 Toulouse, France}
\altaffiltext{10}{Department of Astronomy and Space Sciences, Science Faculty, Istanbul University, Beyaz\i t, 34119 Istanbul, Turkey}
\altaffiltext{11}{Istanbul University Observatory Research and Application Center, Beyaz\i t, 34119 Istanbul, Turkey}
\altaffiltext{12}{Eureka Scientific, Inc., 2452 Delmer Street, Oakland, CA 94602, USA}
\altaffiltext{13}{SRON, Netherlands Institute for Space Research, Sorbonnelaan 2, 3584 CA Utrecht, The Netherlands}
\altaffiltext{14}{Center for Theoretical Astrophysics and Department of Physics, University of Illinois at Urbana-Champaign, 1110 West Green Street, Urbana, IL 61801-3080, USA}
\altaffiltext{15}{Department of Astronomy, University of Illinois at Urbana-Champaign, 1002 West Green Street, Urbana, IL 61801-3074, USA}
\altaffiltext{16}{Department of Astronomy, University of Michigan, 1085 South University Avenue, Ann Arbor, MI 48109-1107, USA}
\altaffiltext{17}{Space Science Division, Naval Research Laboratory, Washington, DC, 20375-5352, USA}

\email{lkeek@umd.edu}
\begin{abstract}
 Accretion disks around neutron stars regularly undergo sudden strong
irradiation by Type I X-ray bursts powered by unstable thermonuclear
burning on the stellar surface. We investigate the impact on the disk
during one of the first X-ray burst observations with the \emph{Neutron
Star Interior Composition Explorer} (\emph{NICER}) on the International
Space Station. The burst is seen from \source{} during the hard spectral
state. In addition to thermal emission from the neutron star, the
burst spectrum exhibits an excess of soft X-ray photons below 1 keV,
where \emph{NICER}'s sensitivity peaks. We interpret the excess as
a combination of reprocessing by the strongly photoionized disk and
enhancement of the pre-burst persistent flux, possibly due to Poynting
Robertson drag or coronal reprocessing. This is the first such detection
for a short sub-Eddington burst. As these bursts are observed frequently,
\emph{NICER} will be able to study how X-ray bursts affect the disk
and corona for a range of accreting neutron star systems and disk
states.
\end{abstract}


\section{Introduction}

In June 2017 the \emph{Neutron Star Interior Composition Explorer}
\citep[NICER;][]{Gendreau2017} was installed on the International
Space Station. Among its first observations were two Type I X-ray
bursts from the low-mass X-ray binary \object[Aql X-1]{Aquila~X-1}.
X-ray bursts are known from over $100$ such systems in our Galaxy,
where hydrogen- and helium-rich material is accreted from a companion
star onto a neutron star \citep[for a recent review, see][]{Galloway2017Review}.
Runaway thermonuclear fusion of the accreted matter powers a brief
(typically $10-100\,\mathrm{s}$) X-ray flash during which the neutron
star outshines the inner regions of the accretion disk. Sudden strong
irradiation can have a multitude of effects on the disk \citep{Ballantyne2005},
but it has been challenging to detect changes in the accretion environment,
because the majority of burst observations have constrained only the
thermal emission from the neutron star \citep[e.g.,][]{swank1977,Galloway2008catalog}.

Most burst observations have been performed with instruments that
are sensitive to photon energies above $\sim3\,\mathrm{keV}$, such
as the Proportional Counter Array \citep[PCA;][]{Jahoda2006} on the
\emph{Rossi X-ray Timing Explorer} \citep[RXTE;][]{Bradt1993}. The
burst spectra are usually fit with a thermal (blackbody) component
in addition to a constant ``persistent'' component. The latter describes
the X-ray emission from accretion processes as measured outside of
the burst, and is assumed to remain unchanged during the burst. Deviations
from the burst spectral model are found when considering a large sample
of observations with \emph{RXTE}/PCA \citep{Worpel2013,Worpel2015},
which may indicate reprocessing of the burst emission or enhancement
of the accretion flow due to the burst's radiation drag on the disk
(Poynting-Robertson drag; \citealp[e.g., ][]{Walker1992,Miller1993,Lamb1995}),
and a deficit of photons at $>30\,\mathrm{keV}$ during the bursts
suggests coronal cooling \citep{Maccarone2003,Chen2012,Chen2013,Ji2014,Kajava2016}.
Furthermore, hours-long superbursts exhibit an iron emission line
and absorption edge produced by reprocessing of the burst by the inner
disk \citep[e.g.,][]{Ballantyne2004models}. Often collectively referred
to as ``reflection'', reprocessing involves both scattering and
absorption/re-emission by the disk. The shape of the reflection spectrum
depends on the ionization of the metals in the disk and on its inner
radius, $R_{\mathrm{in}}$, because relativistic Doppler broadening
is stronger close to the neutron star. The two superbursts seen by
\emph{RXTE}/PCA strongly ionized the disk and temporarily disrupted
the inner disk \citep{Ballantyne2004,Keek2014sb2}. 

Most bursts are too short to enable detection of the iron line. Further
reflection features are predicted in the soft X-ray band below $3\,\mathrm{keV}$,
including a multitude of emission lines on top of a free-free continuum
\citep{Ballantyne2004models}. Burst reflection may, therefore, explain
the soft excess over a blackbody detected during a bright burst observed
with both \emph{Chandra} and \emph{RXTE}/PCA \citep{Zand2013} and
two long bursts seen with the \emph{Swift} X-Ray Telescope \citep{Degenaar2013,Keek2017}.
Moreover, an increase of the persistent emission may also contribute
to the soft excess.

\emph{\facility{NICER}} combines a $0.2-12\,\mathrm{keV}$ passband
with high throughput, and provides a substantially larger effective
area around $1\,\mathrm{keV}$ than previous missions. It offers the
exciting opportunity to study reflection and other signatures of burst-disk
interaction even during short bursts. In this Letter we investigate
one of \emph{NICER}'s first X-ray burst observations: a bright burst
from \source{}. This source exhibits frequent accretion outbursts
during which X-ray bursts have been observed \citep{Koyama1981},
and disk reflection has been detected in the persistent emission \citep{King2016,Ludlam2017}.
After describing the \emph{NICER} observations of \source{} (Section~\ref{sec:Observations-and-Spectral}),
we perform a detailed analysis of the soft excess in the burst spectrum
(Section~\ref{sec:pers_spectra}). We discuss how an enhanced persistent
component and disk reflection contribute (Section~\ref{sec:Discussion}),
and conclude that \emph{NICER}'s ability to detect burst-disk interaction
in short bursts enables investigations for a wide range of sources
and spectral states.

\section{Observations}

\label{sec:Observations-and-Spectral} 

\emph{NICER}'s X-ray Timing Instrument \citep[XTI;][]{Gendreau2016}
consists of $56$ co-aligned X-ray concentrator optics each paired
with a silicon-drift detector \citep{Prigozhin2012}. The XTI provides
a peak effective collecting area of $1900\,\mathrm{cm^{2}}$ and a
$<100\,\mathrm{eV}$ energy resolution at $1.5\,\mathrm{keV}$. In
the interval 2017 June~20 -- July~3, \emph{NICER} collected with
$52$ functioning detectors a total good exposure of $51\,\mathrm{ks}$
on \source{} during a hard-state accretion outburst. Two Type-I bursts
were observed: one in ObsID 0050340108 at MJD 57936.58042 with a
peak rate of $2248\,\mathrm{c\,s^{-1}}$, and another in ObsID 0050340109
at MJD 57937.61102 peaking at $3228\,\mathrm{c\,s^{-1}}$. Neither
burst shows significant oscillations near the neutron star's $550\,\mathrm{Hz}$
spin frequency \citep{Zhang1998}. 

\begin{figure}
\includegraphics{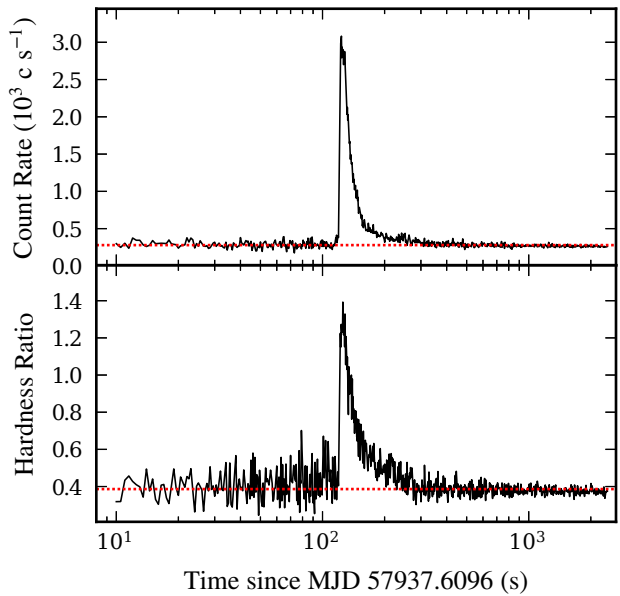}

\caption{\emph{\label{fig:lcv_aql}}(Top) Light curve of the \emph{NICER} pointing
with the burst on 2017 July 3. Initially, the plotted resolution is
$0.5$~s, and $500$ logarithmically spaced bins are employed after
$150\,\mathrm{s}$. (Bottom) Hardness ratio of count rate with $E>2.5\,\mathrm{keV}$
to $E<2.5\,\mathrm{keV}$. The dotted lines indicate mean values over
the first $100\,\mathrm{s}$. }
\end{figure}

\begin{figure}
\includegraphics{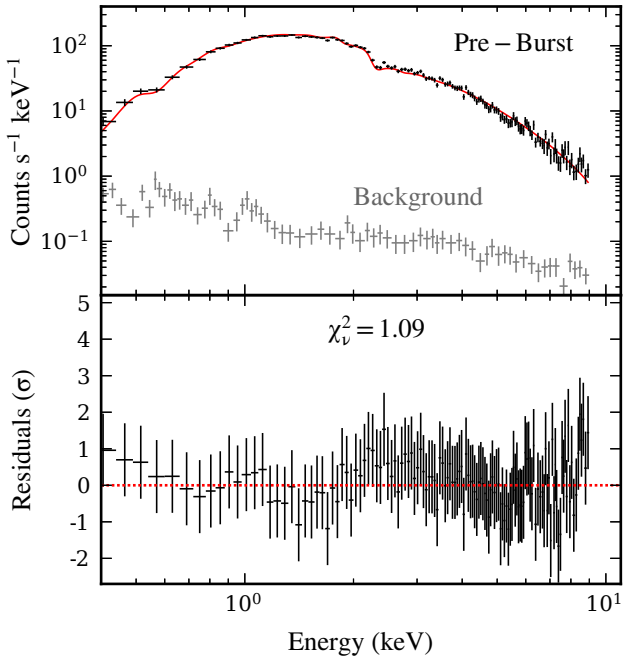}

\caption{\emph{\label{fig:preburst_fit}} Spectral fit to the pre-burst spectrum
as a function of energy, $E$. Top: pre-burst spectrum and best-fitting
absorbed bremsstrahlung model (solid line). The background (rebinned)
is small with respect to the source at all energies, and has been
subtracted from the shown pre-burst spectrum. The absorption edge
near $2.3\,\mathrm{keV}$ is instrumental. Bottom: fit residuals and
the goodness of fit, $\chi_{\nu}^{2}$.}
\end{figure}

In this Letter we analyze the brighter of the two bursts (Figure~\ref{fig:lcv_aql};
the other suffers from a high particle background). During its observation
the instrument pointing was accurate and stable. The ISS was on the
nightside of the Earth, and the Moon was not near the pointing direction,
such that optical loading effects were not significant. The ISS was
not near the high particle background region of the South Atlantic
Anomaly. By virtue of \emph{NICER}'s modularity, dead time and pile-up
are not an issue even at the burst peak. We process and analyze the
data using \textsc{Heasoft} version 6.22.1, \textsc{Nicerdas} 2017-09-06\_V002,
\textsc{Xspec} 12.9.1p \citep{Arnaud1996}, and version 0.06 of the
\emph{NICER} response files. Gain is calibrated separately for each
detector. As a measure of the cosmic and instrument background, we
create a spectrum from a $1117\,\mathrm{s}$ blank-field observation
of \emph{RXTE} background region 5 \citep{Jahoda2006}, which was
also obtained at night. The count rate as a function of energy is
$<1\,\mathrm{c\,s^{-1}keV^{-1}}$, such that our observations of \source{}
are strongly source dominated at all energies (Figure~\ref{fig:preburst_fit}).
In the source spectra we group neighboring spectral bins with fewer
than $15$ counts, and in our Figures we rebin the spectra to a bin
width of at least $50\,\mathrm{eV}$ (\emph{NICER} data oversample
the detector resolution).

\section{Results}

\label{sec:pers_spectra}

\begin{figure*}
\begin{centering}
\includegraphics{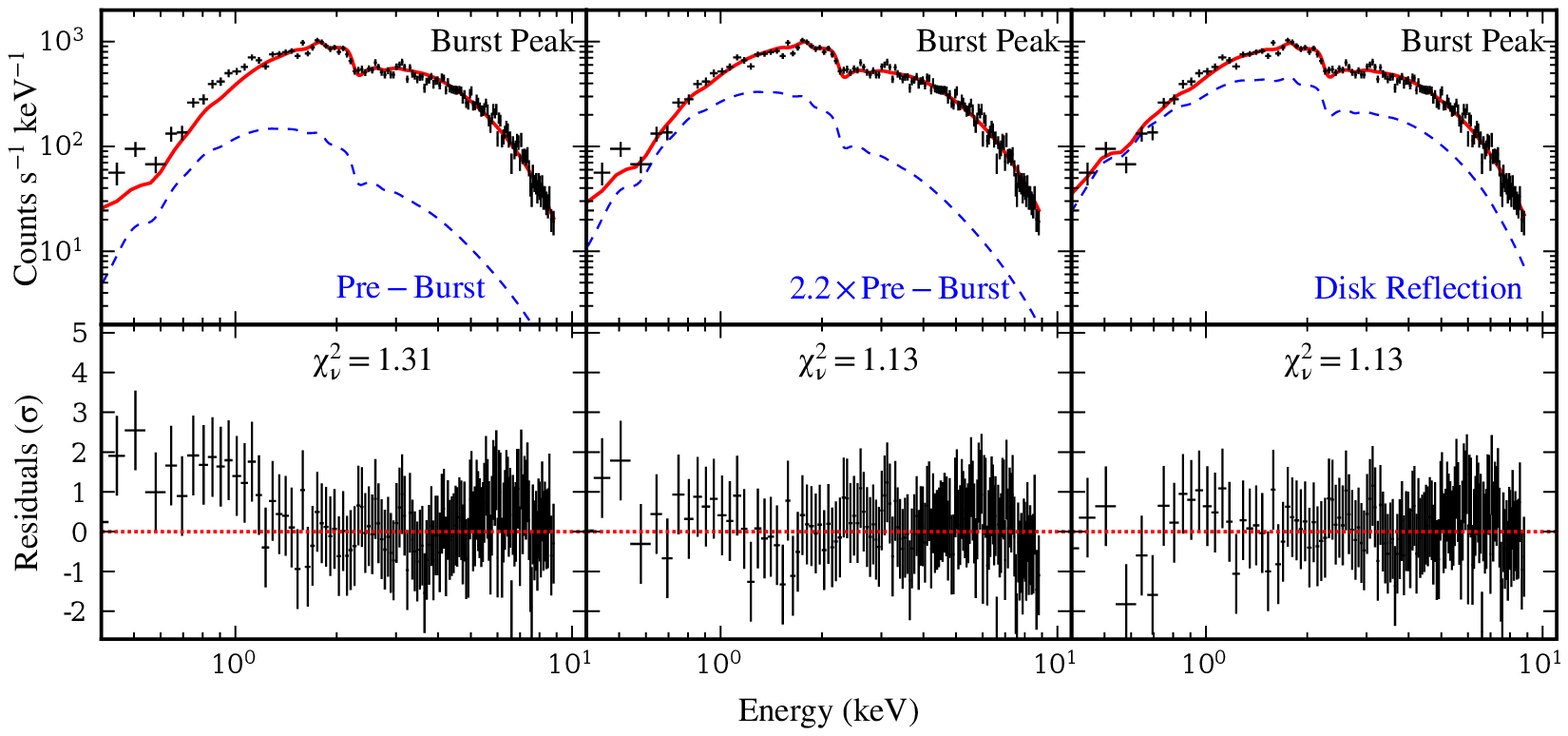}
\par\end{centering}
\caption{\emph{\label{fig:fit_peak}}Spectral fits to burst peak (top panels;
solid line is the best-fit model and dashed line illustrates one model
component) and the fit residuals with the goodness of fit, $\chi_{\nu}^{2}$
(bottom panels). Left: the commonly used spectral model with a blackbody
and fixed pre-burst spectrum leaves a substantial soft excess. Middle:
an increase in the normalization of the pre-burst component fits the
excess. Right: alternatively, a disk reflection component produces
a similar improvement.}
\end{figure*}

We analyze the spectra of the persistent emission prior to the burst
and of the burst itself, looking for signatures of the impact of the
burst on the accretion environment.

\subsection{Pre-burst Emission}

\label{subsec:Pre-burst-emission}

The burst occurred near the pointing's onset, and the source does
not exhibit variability outside the burst (Figure~\ref{fig:lcv_aql}).
We analyze the pre-burst persistent emission from a $102\,\mathrm{s}$
interval. We limit the analysis to the $0.3-9.0$~keV band to avoid
potential noise at both ends of the passband. The spectrum is well
described by a thermal bremsstrahlung model (\texttt{bremss} in \textsc{Xspec};
e.g., \citealt{Czerny1987}). More complex models \citep[e.g.,][]{Ludlam2017}
require a longer exposure to constrain their parameters, whereas our
primary need is a simple description of the persistent spectrum. Interstellar
absorption is modeled using the Tübingen-Boulder model (\texttt{TBabs})
with abundances from \citet{Wilms2000}. With a goodness of fit of
$\chi_{\nu}^{2}=1.09$ for $\nu=514$ degrees of freedom, the best-fitting
plasma temperature is $\mathrm{k}T=31\pm4\,\mathrm{keV}$, the unabsorbed
in-band flux is $(1.66\pm0.02)\times10^{-9}\,\mathrm{erg\,s^{-1}\,cm^{-2}}$,
and the absorption column is $N_{\mathrm{H}}=(5.39\pm0.09)\times10^{21}\,\mathrm{cm^{-2}}$
($1\sigma$ uncertainties). 

Extrapolating the bremsstrahlung model over the $0.001-100$~keV
range, we find an unabsorbed bolometric flux of $F_{\mathrm{pre-burst}}=(4.3\pm0.3)\times10^{-9}\,\mathrm{erg\,s^{-1}\,cm^{-2}}$.
We compare it to the peak fluxes of Eddington-limited bursts from
\source{} observed with \emph{RXTE}/PCA: $F_{\mathrm{Edd}}=(1.0\pm0.2)\times10^{-7}\,\mathrm{erg\,s^{-1}\,cm^{-2}}$
\citep{Worpel2015}. The pre-burst flux level is, therefore, $\sim5\%\,F_{\mathrm{Edd}}$.

The value of $N_{\mathrm{H}}$ is well within the range derived from
observations of the source with the \emph{XMM-Newton}, \emph{Chandra},
and \emph{Swift} observatories \citep[e.g.,][]{Campana2014}. $N_{\mathrm{H}}$
measurements from radio and infrared maps of Galactic hydrogen\footnote{See \url{http://www.swift.ac.uk/analysis/nhtot/}}
\citep{Schlegel1998,Kalberla2005,Willingale2013} find within $1^{\circ}$
of the source $N_{\mathrm{H}}=4.30\times10^{21}\,\mathrm{cm^{-2}}$,
which is $20\%$ lower than our value. Because the difference is modest,
we use our value in the burst analysis for consistency.

\subsection{Burst Peak}

\label{subsec:Burst-emission}

We extract a spectrum around the time when the flux peaks (Figure~\ref{fig:fit_peak})
during a $6\,\mathrm{s}$ interval (starting at $2.3\,\mathrm{s}$
in Figure~\ref{fig:fa_fit}). First we employ the commonly used spectral
model for bursts: we keep the parameters of the pre-burst spectrum
fixed, and we add an absorbed blackbody (\texttt{bbodyrad}) component
to model the thermal emission from the burst. When left free, the
best-fitting value of $N_{\mathrm{H}}$ is substantially smaller than
both the pre-burst fit and the Galactic hydrogen maps indicate. Similar
to \citet{Keek2017}, we fix $N_{\mathrm{H}}$ to the pre-burst value:
the fit yields $\chi_{\nu}^{2}=1.31$ ($\nu=560$), and a substantial
soft excess is visible in the fit residuals below $E\lesssim1\,\mathrm{keV}$
(Figure~\ref{fig:fit_peak} left). At $0.8\,\mathrm{keV}$, the observed
count rate is $\sim2$ times the value of the best-fitting blackbody
model, and the excess is $\sim500$ times the background.

We investigate two interpretations of the soft excess. Following \citet{Worpel2013}
we include a multiplication factor, $f_{\mathrm{a}}$, for the normalization
of the bremsstrahlung component. We find a best-fitting value of $f_{\mathrm{a}}=2.24\pm0.13$.
The fit is improved ($\chi_{\nu}^{2}=1.13$, $\nu=559$), and the
soft-excess is largely removed from the residuals (Figure~\ref{fig:fit_peak}
middle). Comparing this fit to the previous fit with an F-test indicates
a significant improvement with a null-hypothesis probability of $5\times10^{-20}$.

Alternatively, the excess may result from reprocessing by the disk.
We employ the burst reflection model that was successfully applied
to the two \emph{RXTE}/PCA superbursts \citep{Ballantyne2004models,Ballantyne2004,Keek2014sb2},
which consists of a table of detailed reflection spectra calculated
for blackbody illumination of a (in this case, solar-composition)
disk. A reflection component is added to our spectral model, and the
absorption and bremsstrahlung parameters are kept fixed. Relativistic
Doppler broadening of the reflection component is modeled with the
\texttt{rdblur} convolution model \citep{Fabian1989}, using an emissivity
profile that drops off with the third power of the disk radius, and
assuming a disk inclination angle of $20^{\circ}$ \citep{King2016}.
We find a similar significant improvement in the fit as with the $f_{\mathrm{a}}$
model: $\chi_{\nu}^{2}=1.13$ ($\nu=557$; Figure~\ref{fig:fit_peak}
right). The reflection fraction (the flux ratio of the reflection
to blackbody component) is $f_{\mathrm{refl}}=0.37\pm0.05$. As no
discrete features such as lines are visible, the fit prefers the largest
values of the ionization parameter in our table model, $\log\xi=3.66_{-0.11}^{+p}$,
where ``$p$'' indicates that the search for the $1\sigma$ confidence
region is pegged at the table boundary of $\log\xi=3.75$. For similar
reasons, the fit prefers the strongest broadening, which is given
by an inner disk radius of $R_{\mathrm{in}}=6\,R_{\mathrm{g}}$ ($R_{\mathrm{g}}=GM/c^{2}$
is the gravitational radius), but this parameter is not strongly constrained.

\subsection{Time-resolved Spectroscopy}

\label{subsec:Time-Resolved-Spectroscopy}

\begin{figure}
\includegraphics{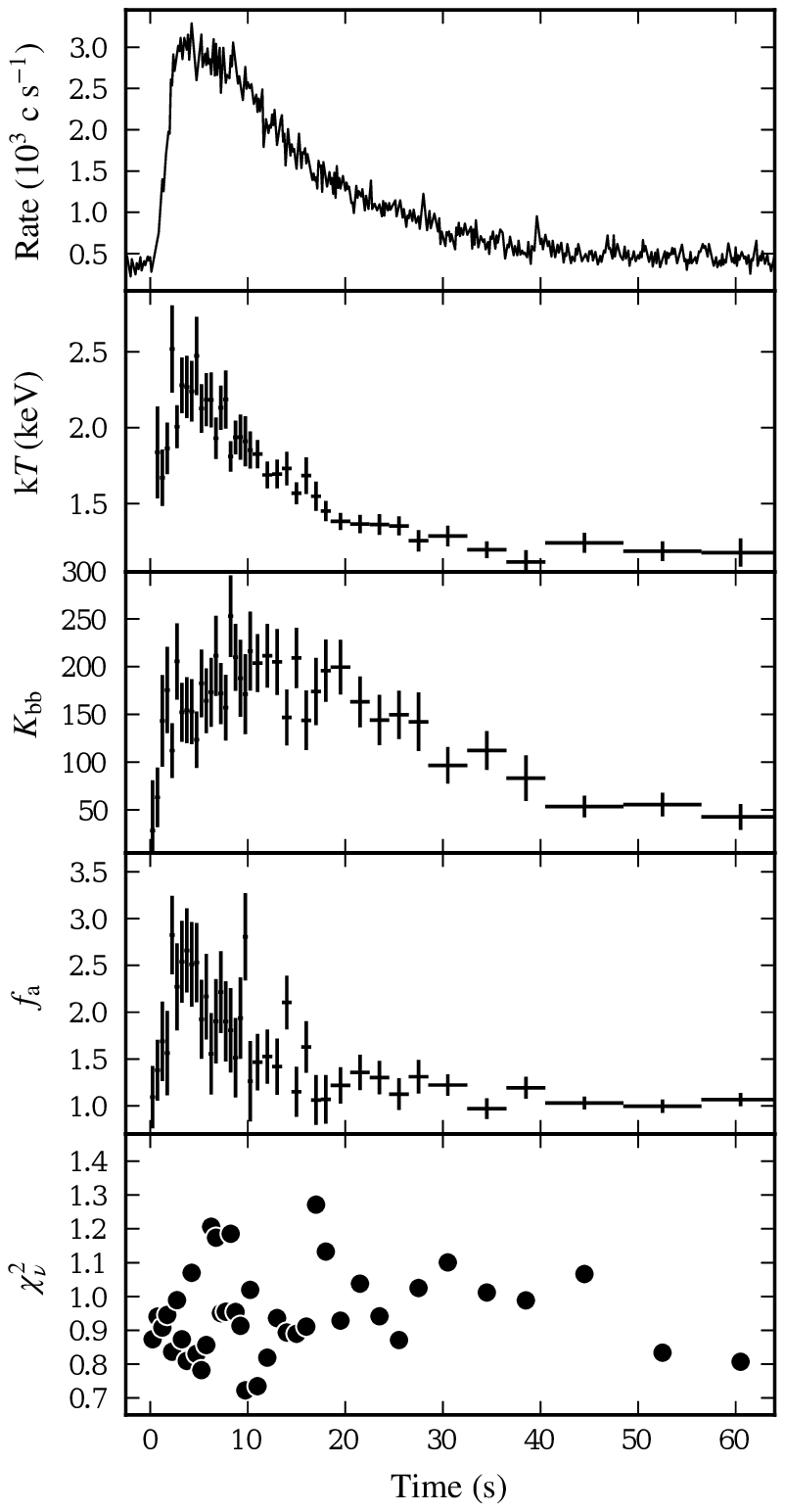}

\caption{\emph{\label{fig:fa_fit}}Time-resolved spectroscopy with a model
that includes a scaled pre-burst component. Top panel: count rate
in the full passband at $1/8\,\mathrm{s}$ resolution. Other panels
show the best-fit values for the blackbody temperature, $\mathrm{k}T$,
and normalization, $K_{\mathrm{bb}}$, as well as the scaling factor
of the pre-burst component, $f_{\mathrm{a}}$. Horizontal bars indicate
the width of the time bins, and vertical bars are the $1\sigma$ uncertainties.
The bottom panel presents the goodness of fit per degree of freedom,
$\chi_{\nu}^{2}$.}
\end{figure}
 We analyzed a $6\,\mathrm{s}$ time interval around the peak, but
the spectral parameters evolve on shorter time scales, requiring time-resolved
spectroscopy. We divide the burst into intervals of $0.5\,\mathrm{s}$
at the burst onset, and after the peak we double the duration each
time the count rate drops by another factor of $\sqrt{2}$, such that
we have similar statistics throughout the burst. We analyze the first
minute of the burst where the parameters of two spectral components
can be constrained, although the tail of the burst is detected for
another $\sim120\,\mathrm{s}$ due to \emph{NICER}'s soft-band sensitivity
to declining temperatures (Figure~\ref{fig:lcv_aql}).

\begin{figure}
\includegraphics{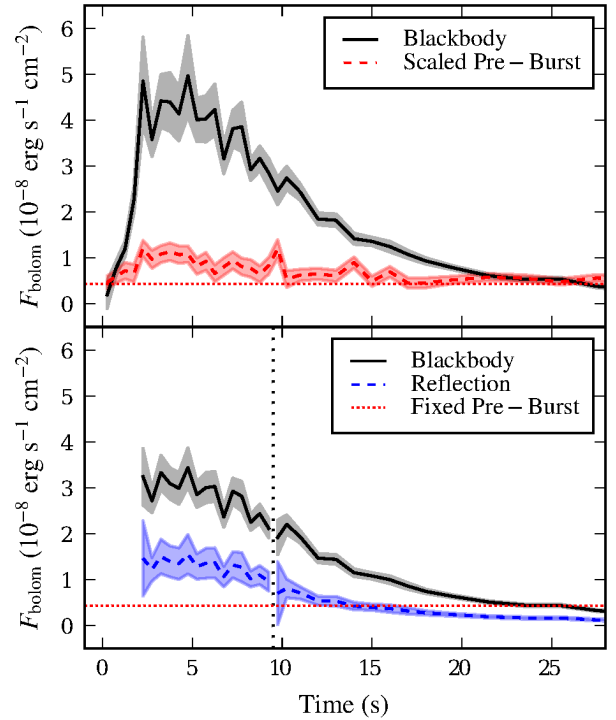}

\caption{\emph{\label{fig:time_resolved}}Bolometric unabsorbed flux from time-resolved
spectroscopy, where the shaded bands indicate the $1\sigma$ error
regions. Top: fit with scaled pre-burst component shows this part
to increase during the burst (dotted line indicates the pre-burst
flux level). Bottom: alternative fit with reflection model. The vertical
dotted line demarcates the two fitted intervals of the peak and tail.}
\end{figure}

First we fit the $f_{\mathrm{a}}$ model. $f_{\mathrm{a}}$ increases
at the burst onset to a maximum, and returns to $1$ in the tail (Figure~\ref{fig:fa_fit}).
In seven bins around the peak, the weighted mean is $f_{\mathrm{a}}=2.5\pm0.2$.
We use the \textsc{Xspec} model \texttt{cflux} to determine the unabsorbed
bolometric flux of the spectral components (Figure~\ref{fig:time_resolved}
top). The flux of the scaled pre-burst component, $F_{\mathrm{bremss}}$,
follows the blackbody flux, $F_{\mathrm{bb}}$: a linear fit to the
first $15\,\mathrm{s}$ of the burst yields $F_{\mathrm{bremss}}=F_{\mathrm{pre-burst}}+(0.128\pm0.012)F_{\mathrm{bb}}$
($\chi_{\nu}^{2}=0.83$, $\nu=24$), where $F_{\mathrm{pre-burst}}$
is the persistent flux from Section~\ref{subsec:Pre-burst-emission}. 

Next we repeat the time-resolved fits with the reflection model. This
model includes more parameters, which are hard to constrain within
short time intervals. Because $R_{\mathrm{in}}$ was weakly constrained
in Section~\ref{subsec:Burst-emission}, we fix its value to $R_{\mathrm{in}}=15\,R_{\mathrm{g}}$,
inferred from reflection spectroscopy of the persistent emission in
the soft state of \source{} \citep{King2016}. An updated analysis
of the soft state \citep{Ludlam2017} and a study of the hard state
\citep{Sakurai2012} reveal similar $R_{\mathrm{in}}$, and our burst
results are insensitive to the differences. A preliminary investigation
of the persistent emission from all \emph{NICER} data on \source{}
shows a broad Fe line similar in shape to that seen with \emph{NuSTAR}
during the soft state \citep{Ludlam2017}, supporting our choice of
$R_{\mathrm{in}}$. Furthermore, we limit the fit to the spectra in
an $8\,\mathrm{s}$ interval around the peak (starting at $2\,\mathrm{s}$
in Figure~\ref{fig:fa_fit}), where we can reasonably expect $\log\xi$
to be large. We fit all spectra in that interval simultaneously, assuming
$\log\xi$ and $f_{\mathrm{refl}}$ to be the same everywhere, and
$\mathrm{k}T$ is fit for each spectrum. We find $\log\xi=3.75_{-0.2}^{+p}$
and $f_{\mathrm{refl}}=0.45\pm0.08$ ($\chi_{\nu}^{2}=0.93$ with
$\nu=1244$). $\mathrm{k}T$ is consistent within $1\sigma$ with
the values from the fit with the $f_{\mathrm{a}}$ model (Figure~\ref{fig:fa_fit}).
Immediately following this time interval, we repeat this exercise
for the tail of the burst, obtaining $\log\xi=3.75_{-0.2}^{+p}$ and
$f_{\mathrm{refl}}=0.37\pm0.11$ ($\chi_{\nu}^{2}=0.95$ with $\nu=1731$),
which are consistent with the values around the peak. The blackbody
flux is lower than for the $f_{\mathrm{a}}$ model (Figure~\ref{fig:time_resolved}),
because the reflection model also contributes to the thermal continuum.

In sevens bins around the peak, the weighted mean of the bolometric
unabsorbed blackbody flux is $F_{\mathrm{bb}}=(4.1\pm0.2)\times10^{-8}\,\mathrm{erg\,s^{-1}\,cm^{-2}}$
for the $f_{\mathrm{a}}$ model and $F_{\mathrm{bb}}=(2.7\pm0.4)\times10^{-8}\,\mathrm{erg\,s^{-1}\,cm^{-2}}$
for the reflection model ($F=(4.2\pm0.2)\times10^{-8}\,\mathrm{erg\,s^{-1}\,cm^{-2}}$
including the reflection component), which is $\sim40\%\,F_{\mathrm{Edd}}$
(Section~\ref{subsec:Pre-burst-emission}). The bolometric flux at
the peak of all components combined (including pre-burst for the reflection
fit) is $F_{\mathrm{total}}=(5.1\pm0.3)\times10^{-8}\,\mathrm{erg\,s^{-1}\,cm^{-2}}$
for the $f_{\mathrm{a}}$ model and $F_{\mathrm{total}}=(4.8\pm0.2)\times10^{-8}\,\mathrm{erg\,s^{-1}\,cm^{-2}}$
for the reflection model, which are consistent within $1\sigma$.

\section{Discussion}

\label{sec:Discussion}

We find that both an enhanced persistent component and disk reflection
can explain the soft excess detected by \emph{NICER} in a burst from
\source{}. Other interpretations may be possible. For example, free-free
absorption in the neutron star atmosphere could produce a soft excess
\citep[e.g.,][]{Suleimanov2012}. However, we find that fits with
atmosphere models are unable to explain the full soft excess, whereas
reflection and enhanced persistent emission were found to be important
for the interpretation of, e.g., the superbursts seen with \emph{RXTE}/PCA
\citep{Ballantyne2004,Keek2014sb2}. Here we discuss how these two
components can provide a consistent physical picture of the impact
of the burst.

\subsection{Enhanced Persistent Emission}

The $f_{\mathrm{a}}$ model scales the persistent flux prior to the
burst. The peak value of $f_{\mathrm{a}}=2.5\pm0.2$ is typical for
bursts without photospheric expansion at a similar persistent flux
\citep{Worpel2015}. \citet{Worpel2015} used \emph{RXTE}/PCA spectra,
which do not cover the soft band $E\lesssim3\,\mathrm{keV}$, and
it is therefore interesting that we obtain a roughly similar value.
We find that the increase in the persistent flux is proportional to
the blackbody flux, suggesting that the increase is caused by burst
irradiation. It is possible that radiation drag enhances accretion
during the burst \citep{Worpel2013}. Alternatively, the soft excess
may be produced by reprocessing of the burst flux in an optically
thin medium such as the corona. \emph{RXTE}/PCA observations of similar
bursts from \source{} in the hard state exhibit a substantial flux
decrease in the $40-50\,\mathrm{keV}$ band, possibly caused by coronal
cooling induced by the burst \citep{Chen2013}. The simplistic $f_{\mathrm{a}}$
model does not probe this temperature evolution, and the application
of physically better motivated models requires broad-band observations
with \emph{NICER} and \emph{NuSTAR} or \emph{ASTROSAT}.

\subsection{Disk Reflection}

The soft excess could also be produced by reprocessing on the disk
\citep{Ballantyne2004models}. For a highly ionized disk, the fluorescent
Fe~K$\alpha$ line from reflection is challenging to detect during
a burst with \emph{NICER}, whereas the soft excess is highly significant
\citep{Keek2016reflsim}. For an inclination angle of $20^{\circ}$
\citep{King2016} the expected reflection fraction is $f_{\mathrm{refl}}=0.52$
for a thin disk that extends to the neutron star \citep{He2016}.
The disk has, however, been observed to truncate at $R_{\mathrm{in}}\simeq15\,R_{\mathrm{g}}$
\citep{King2016,Sakurai2012}. From \citet{He2016} Figure~5 we estimate
that for a $1.4\,M_{\odot}$ and $10\,\mathrm{km}$ radius neutron
star this gap reduces the reflection fraction to $f_{\mathrm{refl}}\simeq0.15$.
We find triple this value, $f_{\mathrm{refl}}=0.45\pm0.08$, for the
burst. If one assumes that the burst does not change the disk geometry,
only $1/3$ of the soft excess is due to reflection, and the rest
could result from an enhancement of the persistent flux. To produce
$f_{\mathrm{refl}}=0.45$, the impact of the burst must have caused
the inner disk to (temporarily) extend close to the neutron star surface.
A similar suggestion was made for the long burst from IGR~J17062$-$6143
\citep{Keek2017}. 

The intermittent presence of dips observed from \source{} may hint
at a larger inclination angle of $\sim75^{\circ}$ \citep{Galloway2016}.
At this angle, $f_{\mathrm{refl}}=0.10$ is expected when the disk
extends to the stellar surface \citep{He2016}, and $f_{\mathrm{refl}}=0.03$
for $R_{\mathrm{in}}\simeq15\,R_{\mathrm{g}}$. Under these assumptions,
reflection contributes only a small part of the soft excess detected
by \emph{NICER}.

Unfortunately, we could not track $R_{\mathrm{in}}$ during the burst
from the reflection signal. This may require\emph{ }the analysis of
a larger sample of bursts, a superburst observation with \emph{NICER},
or a future mission with even larger collecting area such as \emph{STROBE-X
}\citep{WilsonHodge2017}\emph{.}

\section{Conclusions and Outlook}

\label{sec:Conclusions-and-Outlook}

One of \emph{NICER}'s first X-ray burst observations was a sub-Eddington
burst from \source{} in the hard state. The spectrum exhibits a soft
excess over the thermal burst emission, which can be explained by
either enhanced persistent (accretion) emission or disk reflection.
From the known disk truncation radius, we expect at least a third
of the excess to be disk reflection. For reflection to produce all
of the excess, burst irradiation must cause the inner disk to temporarily
move close to the stellar surface. Alternatively, the excess may be
powered by Poynting-Robertson drag or coronal reprocessing. Regardless
of the precise interpretation, this demonstrates that bursts have
a substantial impact on their accretion environment, even in the hard
spectral state which is preferred for neutron star mass-radius measurements
\citep[e.g.,][]{Kajava2014}. Whereas previously this was only detectable
in rare cases or by considering large samples, a preliminary analysis
of \emph{NICER} burst observations finds the soft excess in a number
of short bursts. This will allow us to study the burst-disk interaction
across multiple sources and spectral states, mapping out how bursts
impact different accretion geometries.

\acknowledgements{This work was supported by NASA through the \emph{NICER} mission
and the Astrophysics Explorers Program. It benefited from JINA CEE
(NSF grant PHY-1430152). EMC acknowledges NSF CAREER award AST-1351222.
SG acknowledges CNES. }

\bibliographystyle{apj}
\bibliography{aqlxnicer}

\end{document}